\documentclass[conference, a4paper, 9pt]{IEEEtran}
\usepackage[english]{babel}
\usepackage[artemisia]{textgreek}
\usepackage[T1]{url}
\usepackage[final]{microtype}
\usepackage{amsmath}
\usepackage[romanConstants, boldVectors]{scstuff}
\usepackage{units}
\usepackage{acronym}
\usepackage{cite}
\usepackage[caption=false, subrefformat=parens]{subfig}
\usepackage[pdftex]{graphicx}
\usepackage{booktabs}
\usepackage[inline]{enumitem}
\usepackage[pdftex, unicode, bookmarks=true, bookmarksnumbered=true,
  pdfpagemode={UseOutlines}, plainpages=false, pdfpagelabels=true,
  colorlinks=true, linkcolor={black}, citecolor={black}, 
  urlcolor={black}]{hyperref}
\usepackage{ucs}
\usepackage[utf8x]{inputenc}
\usepackage{autofe}
\usepackage[svgnames]{xcolor}
  
\graphicspath{{./Figures/}}

\renewcommand\normalsize{\fontsize{8.8}{10.56}\selectfont}

\makeatletter
\let\thanks\@IEEESAVECMDthanks
\makeatother

\makeatletter
\newcommand\isacroused[3]{%
  \expandafter\ifx\csname ac@#1\endcsname\AC@used
  #2\else #3\fi}
\makeatother
\newacro{NTF}{Noise Transfer Function}
\newacro{STF}{Signal Transfer Function}
\newacro{DDSM}[DΔΣM]{Digital ΔΣ Modulator}
\newacro{PCM}{Pulse Code Modulation}
\newacro{CMQ}{Classical Model of Quantization}
\newacro{SSB}{Single Side Band}
\newacro{DSB}{Double Side Band}
\newacro{PSD}{Power Spectral Density}
\newacro{DTFT}{Discrete Time Fourier Transform}
\newacro{DTFS}{Discrete Time Fourier Series}
\newacro{LP}{Low Pass}
\newacro{HP}{High Pass}
\newacro{BP}{Band Pass}
\newacro{OSR}{Oversampling Ratio}
\newacro{FIR}{Finite Impulse Response}

\newcommand{\NTF}{\ensuremath{\mathit{NTF}}}
\newcommand{\STF}{\ensuremath{\mathit{STF}}}

\newcommand{\OSR}{\ensuremath{\mathit{OSR}}}

\title{Coding of Stereo Signals by a Single Digital
  \texorpdfstring{ΔΣ}{Delta-Sigma} Modulator} 
\author{%
  \IEEEauthorblockN{Sergio Callegari}
  \IEEEauthorblockA{ ARCES/DEI, University of Bologna, Italy\\
    \href{mailto:sergio.callegari@unibo.it}%
    {\nolinkurl{sergio.callegari@unibo.it}}}%
  \thanks{This is a pre-print version of a paper published in the Proceedings
    of 2013 IEEE International Conference on Electronics, Circuits, and Systems
    (ICECS). Available via DOI
    \href{http://dx.doi.org/10.1109/ICECS.2013.6815483}%
    {10.1109/ICECS.2013.6815483}. Cite
    as:\protect\\[1ex]
    Callegari, S., ``Coding of stereo signals by a single digital ΔΣ
    modulator,'' \emph{IEEE 20th International Conference on Electronics,
      Circuits, and Systems (ICECS 2013),} pp.~589--592, Dec.\@ 2013.
    \protect\\[1ex]
    Copyright © 2013 IEEE. Personal use of this material is permitted. However,
    permission to use this material for any other purposes must be obtained
    from the IEEE by sending a request to
    \href{mailto:pubs-permissions@ieee.org}{pubs-permissions@ieee.org}.
    \protect\\[-2ex]}}

\PrerenderUnicode{©}

\initializeplaintitle[%
  
  \def\texorpdfstring#1#2{#2}]
\hypersetup{pdftitle=\plaintitle, 
  pdfauthor={Sergio Callegari},
  pdfsubject={Stereo Delta-Sigma Modulation},
  pdfkeywords={Stereo signals, Digital Delta-Sigma Modulation}}

\hypersetup{colorlinks=true, 
  linkcolor=DarkBlue,
  citecolor=DarkBlue,
  urlcolor=DarkBlue}

\begin{document}
\maketitle

\begin{abstract}
  The possibility of using a single digital ΔΣ modulator to simultaneously
  encode the two channels of a stereo signal is illustrated. From the modulated
  stream, the two channels can be recovered with minimal processing and no
  cross-talk. Notably, demultiplexing does not affect the sample-depth so that,
  after it, one still has a data stream suitable for directly driving a power
  bridge and convertible into analog by mere low-pass filtering. Furthermore,
  the approach is very flexible and if one channel is unused, it lets the other
  get improved dynamic range and SNR. The approach can take advantage of recent
  techniques for the design of ΔΣ modulators, including methods for
  psychoacoustically optimal distribution of quantization noise. Code is
  available to replicate the proposed examples and as a general computer aided
  design tool.
\end{abstract}
\acresetall

\section{Introduction}
\ac{DDSM} are the widest adopted form of ΔΣ modulators in commercial integrated
circuits \cite{Pamarti:TCAS1-54-3}. Their role is that of information re-coders
that are fed with a stream of high-resolution samples and deliver an equivalent
high-rate low-depth (low-resolution, e.g., binary, ternary) stream.  By
equivalent, it is meant that from the output stream it is possible to recover
the input information with very good approximation by mere linear filtering. In
other words, the coding process assures that the artifacts created by
resolution reduction have negligible energy in the signal band. The process can
be functional to D/A conversion, frequency synthesis \cite{Harris:WET-2003},
switched mode power control \cite{Bizzarri:ISCAS-2012}, and so on.

A particularly interesting application of \ac{DDSM} is in audio systems, where
D/A conversion is associated to amplification. In conventional setups, a
Nyquist rate D/A converter is followed by a smoothing filter and an analog
amplifier, in an arrangement typically characterized by poor power
efficiency. Conversely, by ΔΣ modulation, a high-resolution digital stream (as
coded in a conventional digital media) can be converted in commands for a
switched mode power bridge.  Power amplification can so be achieved by a
switched-mode unit with much better efficiency. The \ac{DDSM} ability to
deliver a low-depth stream is the key to directly driving the bridge, since the
latter can only assume a very limited number of configurations.

In this paper, the possibility of using a single modulator to simultaneously
encode the two channels of a stereo signal is illustrated. The proposal
introduces a multiplexing mechanism where:
\begin{enumerate*}[label=(\roman*), ref=\roman*]
\item multiplexing and demultiplexing are inexpensive;
\item \ac{DDSM} design techniques specific of the audio realm remain
  applicable; and
\item \label{item:preserve} after demultiplexing the properties of the ΔΣ
  streams are preserved.
\end{enumerate*} 
By (\ref{item:preserve}), one means that the data stream obtained after
de-multiplexing is still suitable to directly drive a switched mode power
bridge as if no multiplexing/demultiplexing was involved.

\begin{figure}[t]
  \begin{tightcenter}
    \includegraphics[width=\lw]{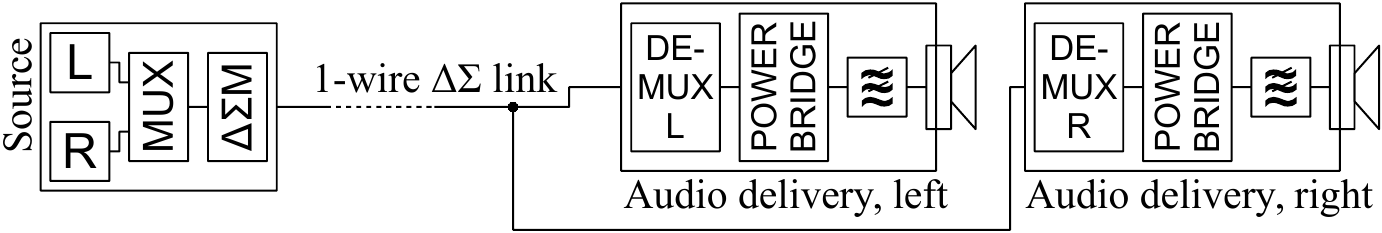}
  \end{tightcenter}
  \caption{Use of a ΔΣ binary stream in a 1-wire link capable of delivering two
    audio channels at once. Connection costs can be reduced if the audio
    actuators are remote from the source.}
  \label{fig:stereo-link}
\end{figure}

The proposed setup saves a \ac{DDSM} with respect to a conventional stereo
arrangement. Yet, this is not a major advantage, since a modulator with more
complex filters is required in exchange. More interesting is the ability to
save connections, particularly when the sound delivery is remote from the
signal source, as in Fig.~\ref{fig:stereo-link}. Even if the loudspeakers need
to be themselves separated from each other, their wiring can be simplified by a
'daisy chain' topology. The advantage increases in environments where passing
wires is difficult or visually unappealing (e.g., domestic) or when specific
connection requirements exist, such as galvanic isolation or optical link,
since in this case a coupling element can be saved too. Obviously, a similar
advantage could also be obtained by conventional digital
multiplexers/demultiplexers in a 2 \acs{DDSM} setup. Yet, the proposed
arrangement is simpler and requires demultiplexing hardware just on one
channel, as it will be shown shortly. Most important, this proposal turns out
to be more flexible. If one channel is not needed, it lets the dynamic range
and SNR be improved on the other.

Even if this paper focuses on audio systems, the same approach can be used
whenever two signals need to be treated at once.  The audio application is
currently targeted both for its economic importance and since it represents an
interesting benchmark. In fact, it imposes strong specifications on cross-talk
and SNR. Furthermore, \acp{DDSM} for the audio domain may include specific
design flows (e.g., for minimizing perceived noise according to psychoacoustic
models \cite{Dunn:JAES-45-4, Callegari:TCAS2-2013}). An appealing feature of
the proposed multiplexing mechanism it its full compatibility with them.  In a
near future, the results presented here will be extended to a larger number of
multiplexed signals and to a wider range of applications (transmission of
multiple sensed data, multiplexing of band-pass signals, storage of multiple
analog waveforms on digital memories, etc.).

\section{Background}
\label{sec:background}
A brief review of ΔΣ modulation is needed to define the notation. ΔΣ coders are
feedback-based nonlinear systems \cite{Schreier:UDSDC-2004} requiring a sample
rate $f_\Phi=\nicefrac{1}{T}$ exceeding twice the width $B$ of the band $\set
B$ of their input signal $u(nT)$ by a large factor known as \ac{OSR}. Their
behavior is typically analyzed relying on a linearized model that replaces the
quantization process responsible for the resolution reduction by the
superposition of a noise signal $\epsilon(nT)$. Following the classical model
of quantization, $\epsilon(nT)$ is assumed to be independent from $u(nT)$,
white and uniformly distributed in value. With this, the modulator behavior
gets characterized through two items: a \ac{STF}, from the input $u(nT)$ to the
output $x(nT)$, and a \ac{NTF}, from $\epsilon(nT)$ to $x(nT)$, so that
$X(z)=\STF(z) U(z)+\NTF(z) E(z)$. Capital letters are here used to indicate the
Laplace transforms of the signals named by the corresponding lower case symbol.

Under common assumptions, a relationship exists between \ac{STF}, \ac{NTF} and
the internal filtering structures inside the modulator such that, once
$\STF(z)$ and $\NTF(z)$ are assigned, the modulator functionality is fully
defined (even if the actual implementation and arrangement of the filters can
vary) \cite{Schreier:UDSDC-2004}. For this reason, the modulator functional
design substantially reduces to a suitable choice of these two transfer
functions \cite{Callegari:TCAS1-2013}. In (re)coding, typically one wants
$u(nT)$ to pass through the modulator unaltered, so that $\STF(z)=1$ or, at
most, $\STF(z)=z^{-d}$ where $d$ is an integer. The \ac{NTF} should then be
arranged to strongly attenuate the quantization noise in $\set B$, so that the
original information can be recovered from $x(nT)$ by merely filtering away all
that is out of $\set B$. To have $\NTF(z)$ highly attenuating in $\set B$, it
must be allowed to amplify elsewhere (so that the net result appears like
moving noise from one frequency region to another)
\cite{Schreier:UDSDC-2004}. This is due to many reasons, not last the
requirement that the modulator feedback loop is non-algebraic. Since the loop
transfer function accounts to $\nicefrac{(1-\NTF(z))}{(\NTF(z))}$, $1-\NTF(z)$
must be \emph{delaying}, and thus $\NTF(z)$ must be biproper and show a unitary
gain when factored in \emph{zero-pole-gain} form. Other commonly constraints
include avoiding the magnitude response of $\NTF(z)$ to peak above a certain
value $\gamma$ depending on the quantizer resolution, following the Lee
stability criterion \cite{Lee:Thesis-1987}. For binary quantizers, $\gamma<2$
and typically $1.5$ is used.

\section{Two way multiplexing in ΔΣ modulation}
\label{sec:multiplexing}
\begin{figure}[t]
  \begin{tightcenter}
    \includegraphics[scale=0.6]{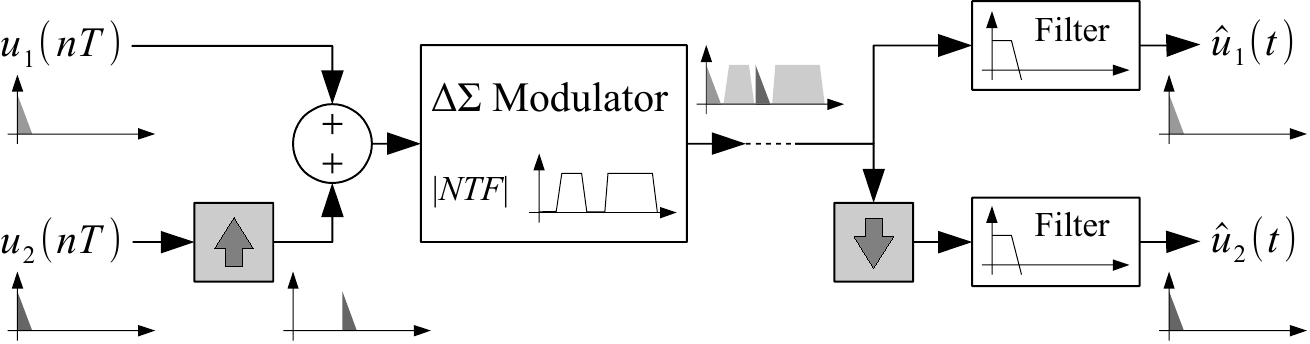}
  \end{tightcenter}
  \caption{Scheme of principle for 2-way frequency division in ΔΣ modulation.}
  \label{fig:upconversion}
\end{figure}

As described in the previous section, the modulator input-output behavior is
approximately equivalent to that of a \emph{linear channel}. Consequently,
multiplexing \ac{LP} signals into a single ΔΣ stream should be practicable by
exploiting \emph{frequency division} and superposition as in
Fig.~\ref{fig:upconversion}. This requires up converting the input signals to
occupy different bands and using an \ac{NTF} with multiple high-attenuation
regions corresponding to them. Then a complementary down conversion can be
applied at the decoding end to restore the original bands, so enabling the
separation of the different streams by filtering. Appealingly, the same filter
that would in any case be used to do signal reconstruction can here double as a
channel separation filter for de-multiplexing.

Notwithstanding apparent simplicity, this approach involves some challenges. In
fact, it requires a modulator whose \ac{NTF} has multiple high-attenuation
zones while respecting the structural constraints summarized in
Sec.~\ref{sec:background}. Furthermore, it involves designing the up- and
down-conversion operations in a way that is acceptable cost-wise and capable of
`passing-through' the low-resolution property of ΔΣ streams after
down-conversion. Without this property, the possibility to directly feed the
down-converted output to a bridge or to transform it into analog by mere
filtering would be hindered. These requirements rule out generic single- and
double-side-band mixers as up-/down-converters since such blocks involve
multiplication by sinusoidal carriers. Not only full-fledged multipliers are
too expensive. They also require high-resolution arithmetic, thus delivering
high-depth outputs unsuitable at the receiving end.

In practice, the sole operation guaranteed to leave the data depth unaltered is
multiplication by $\pm 1$ (since any set of discrete levels balanced around
zero is invariant under this operation). Consequently, down-conversion can only
be based on mixing with carriers $r(nT) \in \{-1,1\} \ \forall
n\in\Nset{Z}$. Interestingly, this mixing is an \emph{involutory} operator,
namely an operator that is its own inverse. In fact, for an arbitrary $u(nT)$
and $r(nT) \in \{-1,1\}$, one has $(u(nT) r(nT)) r(nT) = u(nT) r^2(nT) =
u(nT)\cdot 1 = u(nT)$. This assures that up-conversion, that must necessary be
the inverse of down-conversion, can rely on the very same binary mixing.

Once the restriction to binary mixing is established, what remains to be
evaluated is \emph{how} it can be used to shift spectral occupations in order
to support the multiplexing \emph{and} to guarantee that the \ac{DDSM} has
sufficient space for its quantization noise.  Mixing a generic $u(nT)$ by a
binary periodic $r(nT)$, one gets a signal $y(nT)$ whose \ac{PSD} is
\begin{equation}
  \Psi_y(\hat f) = \int_{0}^{1} \Psi_r(\xi) \Psi_u(\hat f-\xi) d\xi .
  \label{eq:convolution}
\end{equation}
where $\Psi_r(\xi)$ and $\Psi_u(\hat f)$ are the \ac{PSD} of the carrier and
the input signal respectively. In fact, a product in the time domain converts
into a convolution in the frequency domain. Being periodic, $r(nT)$ can be
decomposed in the superposition of multiple complex-exponentials signals as in
\begin{equation}
  r(nT)=\sum_{k=0}^{N-1} r_k \ee^{\ii 2\pi \frac{k n}{N}}
\end{equation}
where $N$ is the period length, $r_k$ is the $k$-th Fourier coefficient and
$r_{N-k}=r^*_k$, with the asterisk indicating complex conjugation. Thus,
\begin{multline}
  \Psi_r(\hat f) = \sum_{n=-\infty}^{\infty} \sum_{k=0}^{N-1} r_k \ee^{\ii 2\pi
    \frac{k n}{N}} \ee^{-\ii 2\pi \hat f n} = \displaybreak[0]\\
  \sum_{k=0}^{N-1} r_k \sum_{i=-\infty}^{\infty} \delta(\hat
  f-\nicefrac{k}{N}+i)
  \label{eq:combs}
\end{multline}
where $\delta(\cdot)$ is the Dirach delta.  When expression~\eqref{eq:combs} is
substituted into~\eqref{eq:convolution}, its argument is restricted within
$[0,1]$, so that only $i=0$ needs to be evaluated and the inner sum
disappears. With this,
\begin{multline}
  \Psi_y(\hat f)=\int_{0}^{1} \Psi_x(\hat f-\xi) \sum_{k=0}^{N-1} r_k
  \delta(\xi-\nicefrac{k}{N}) d\xi = \displaybreak[0]\\
  \sum_{k=0}^{N-1} r_k \Psi_x(\hat f-\nicefrac{k}{N}).
\end{multline}
In other words, $\Psi_y(\hat f)$ is the superposition of $N$ scaled replicas of
the input signal spectrum, shifted by $\nicefrac{1}{N}$ from each other.

If one needs to multiplex two \ac{LP} signals with identical spectral
occupation, the above derivation ensures the possibility of doing so
passing-through one of them without any up-conversion (as in
Fig.~\ref{fig:upconversion}), provided that for the other the up-conversion
involves a binary carrier such that:
\begin{enumerate*}[label=(\roman*),ref=\emph{\alph*}]
  \item it has no dc component (namely $r_0=0$); and
  \item $N<\nicefrac{1}{\hat B}$, with $\hat B=\nicefrac{B}{f_\Phi}$.
\end{enumerate*}
These two properties are sufficient conditions to assure that the up-converted
signal and the passed-through one do not overlap in the frequency domain.

\begin{figure}[t]
  \begin{tightcenter}
    \subfloat[\label{sfig:upconversion-good}]{%
      \shortstack{\includegraphics[scale=0.6]{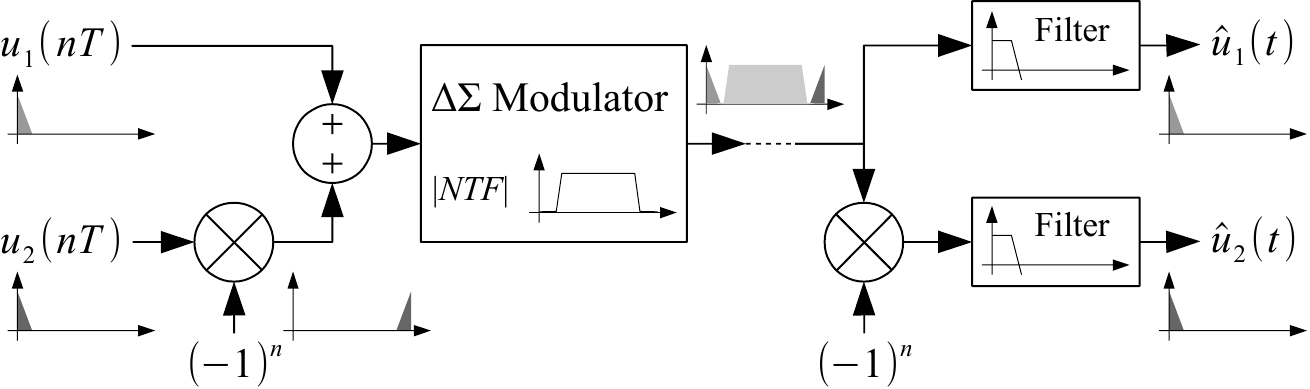}\\[-3ex]}}\\
    \subfloat[\label{sfig:shiftnflip}]{%
      \shortstack{\includegraphics[scale=0.6]{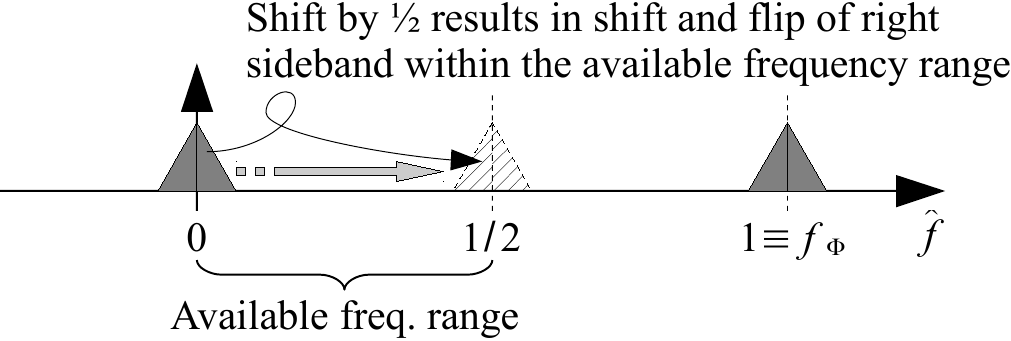}%
        \hspace{2cm}\hnull\\[-3ex]}} 
  \end{tightcenter}
  \caption{Practical architecture for dual channel frequency division in ΔΣ
    modulation \protect\subref{sfig:upconversion-good} and resulting frequency
    conversion \protect\subref{sfig:shiftnflip}.}
  \label{fig:upconversion-good}
\end{figure}

Once this requirement is satisfied, one needs to consider the interactions
between multiplexing and the \ac{DDSM} operation. To make the architecture
comparable to one using two modulators and operating with some \ac{OSR}
indicated as $\OSR$, one needs $f_\Phi=4\,\OSR\, B$, so that the overall data
rate is the same. However, multiplexing makes the modulator operate with an
input waveform whose spectral occupation is made of multiple frequency
intervals each as large as $B$ (due to the spectral replicas), for an overall
bandwidth $B\Ds{eff}=\tilde N B$, where $\tilde N$ is the overall number of
replicas. It is obviously desirable to have $\tilde N$ as low as possible. For
one, the modulator ends up working at an \emph{effective} \ac{OSR} given by
$\OSR\Ds{eff} = \nicefrac{f_\Phi}{(2\,B\Ds{eff})} = \nicefrac{(2\,\OSR)}{\tilde
  N}$, so that its noise performance is necessarily reduced when $\tilde N$ is
large. Secondly, its \ac{NTF} magnitude response needs to have as many valleys
as $\tilde N$, which may require increasing too much the modulator order when
$\tilde N$ is large. However, it is known that $\tilde N$ is certainly bound by
$N$, since it counts the passed-through signal and at most $N-1$ replicas of
the up-converted signal. Thus, picking $N=2$ automatically ensures the smallest
possible $\tilde N$. With this, $r(nT)$ can be set to $(-1)^n$ as in
Fig.~\ref{sfig:upconversion-good}. The resulting up-conversion produces a
single replica shifted by $\hat f=\nicefrac{1}{2}$, so that it basically
converts the \ac{LP} signal to a \ac{HP} one, as shown in
Fig.~\ref{sfig:shiftnflip}.

\section{Modulator design}
\label{sec:modulatordesign}
In the proposed 2-way multiplexed arrangement, the signal entering the
modulator has two components, one \ac{LP} and one \ac{HP}. Consequently, the
\ac{NTF} needs to be \ac{BP}. A suitable \ac{NTF} can be built from a
conventional \ac{NTF} for \ac{LP} signals as follows:
\begin{enumerate*}[label=(\roman*),ref=\emph{\alph*}]
  \item design an \ac{NTF} for the base-band signals, say $\NTF\Ds{LP}(z)$;
  \item obtain the desired \ac{BP} \ac{NTF} as
    $\NTF(z)=\NTF\Ds{LP}(z)\cdot\NTF\Ds{LP}(-z)$.
\end{enumerate*}

Let this procedure be examined in reverse. If $\NTF\Ds{LP}(z)$ is an \ac{HP}
transfer function, then the elementary $z\to -z$ spectral transformation makes
$\NTF\Ds{LP}(-z)$ its equivalent \ac{LP} function. Multiplying $\NTF\Ds{LP}(z)$
and $\NTF\Ds{LP}(-z)$, gets a transfer function that has a high-attenuation
wherever either of its two components has. Furthermore, by symmetry,
$\NTF\Ds{LP}(z)\cdot\NTF\Ds{LP}(-z)$ peaks at $\hat f=\nicefrac{1}{4}$ with a
peak gain approximately given by the squared peak gain of
$\NTF\Ds{LP}(z)$. From this, one finds how $\NTF\Ds{LP}(z)$ needs to be
designed. One must impose a peak gain $\gamma\Ds{LP}=\sqrt{\gamma}$ where
$\gamma$ is the Lee coefficient that would be used in a conventional
non-multiplexed design. This in addition to taking an $\ac{OSR}$ given by
$\OSR\Ds{LP}=2\,\OSR$ (i.e., twice the value that would have been used in a
non-multiplexed setup), following the $f_\Phi$ choice illustrated before.

This design procedure has a major asset in being fully based on the design of
\ac{LP} modulators. Thus, it lets any specific design strategy devised for
\ac{LP} modulators be automatically portable to the 2-way multiplexed
arrangement. For instance, suppose that one wants to base the design on the
\texttt{synthesizeNTF} design strategy proposed by Schreier
\cite{Schreier:UDSDC-2004}. This is implemented into a function
\texttt{synthesizeNTF(order, OSR, opt, H\_inf)}, where \texttt{order} is the
modulator order, \texttt{OSR} is the \ac{OSR}, \texttt{opt} is a flag
controlling some optimization modes, and \texttt{H\_inf} is $\gamma$. Assuming
that one would invoke it as \texttt{synthesizeNTF(order, OSR, opt, H\_inf)} for
a conventional modulator, then to design $\NTF\Ds{LP}(z)$ it is sufficient to
call \texttt{synthesizeNTF(order, OSR/2, opt, sqrt(H\_inf))}, leading to a
final \ac{BP} $NTF(z)$ with an order twice as large as \texttt{order}. All the
optimizations provided by \texttt{synthesizeNTF} will be automatically
present. Similarly, suppose that one has a function capable of designing
psychoacoustically optimal \acp{NTF} (as in \cite{Dunn:JAES-45-4} or
\cite{Callegari:TCAS2-2013}), based on a required order, an \ac{OSR} or
$f_\Phi$ specification, a Lee coefficient value, and possibly other
parameters. Again, it can be used for $\NTF\Ds{LP}(z)$, just remembering to
double the \ac{OSR} (or $f_\Phi$) and to take the square root of the Lee
coefficient with respect to an equivalent non-multiplexed design.

\subsection{Expected performance}
\begin{figure}[t]
  \vspace{-1.5ex}
  \begin{tightcenter}
    \subfloat[\label{sfig:scaling-osr}]{%
      \shortstack{\includegraphics[scale=0.55]{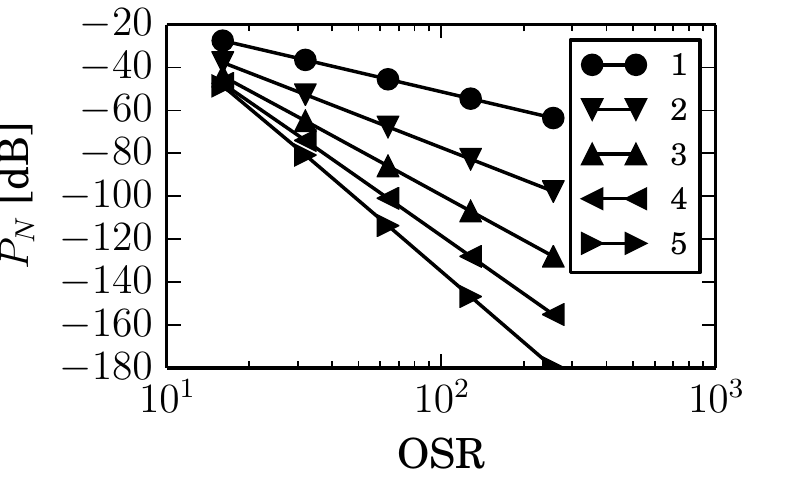}\\[-1ex]}}\hfill
    \subfloat[\label{sfig:scaling-gamma}]{%
      \shortstack{\includegraphics[scale=0.55]{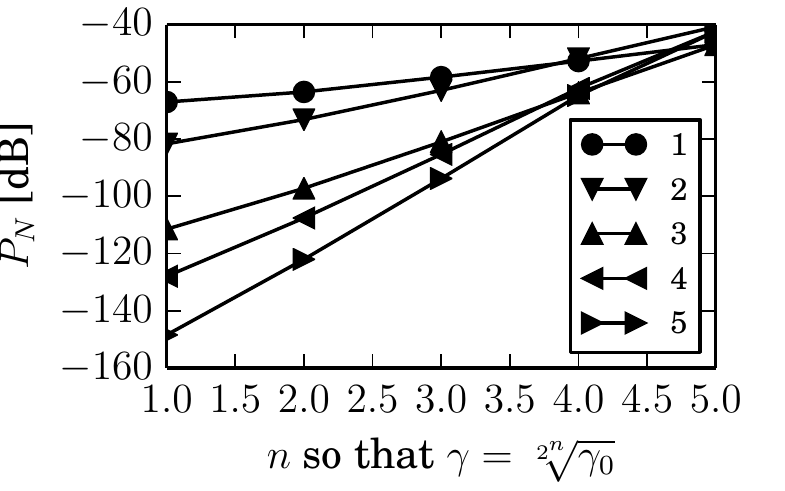}\\[-1ex]}}%
  \end{tightcenter}
  \caption{Effects of rising the \ac{OSR} or reducing the Lee coefficient on
    the modulator noise floor. In \protect\subref{sfig:scaling-osr}, effect of
    changing the \ac{OSR}. In \protect\subref{sfig:scaling-gamma}, effect of
    changing $\gamma$. In this case, $\gamma$ is changed by taking subsequent
    square roots of a reference value $\gamma_0$, so that the noise floor
    worsening corresponding to each square root can be plotted. Tests run for
    multiple modulator orders, using Schreier's \texttt{synthesizeNTF}
    \cite{Schreier:UDSDC-2004} (in \protect\subref{sfig:scaling-osr},
    $\gamma=1.5$; in \protect\subref{sfig:scaling-osr}, the \ac{OSR} is 64 and
    $\gamma_0 = 2.25$.}
\end{figure}

Being ΔΣ modulators strongly non-linear objects, an accurate performance
evaluation can only be based on simulation in actual operating conditions. Yet,
some estimation of the noise-floor and SNR, is still possible by relying on the
linearized model to compute the in-band power level due to quantization noise
as in
\begin{equation}
  P_N=\sigma^2_\epsilon \cdot 2 \int_{0}^{\hat B} \abs{\NTF\left(\ee^{\ii 2\pi \hat
        f}\right)}^2 d\hat f
\end{equation}
where $\sigma^2_\epsilon$ is the power of $\epsilon(nT)$
\cite{Schreier:UDSDC-2004, Callegari:TCAS1-2013}.  In typical cases, $P_N$
scales with the \ac{OSR} in a rather predictable way, as shown in
Fig.~\ref{sfig:scaling-osr}. Specifically, doubling the \ac{OSR} improves the
noise floor by approximately $\unit[3]{dB} + \unit[6]{dB} \cdot
(\text{modulator order})$, which is an expected result
\cite{Schreier:UDSDC-2004}. Furthermore, it can also be empirically found that
$P_N$ scales quite regularly with $\gamma$, as shown in
Fig.~\ref{sfig:scaling-gamma}. Whenever $\gamma$ is reduced by taking its
square root, the noise floor is worsened by approximately $\unit[-1]{dB} +
\unit[6]{dB} \cdot (\text{modulator order})$.

This means that applying the 2-way multiplexing technique, which implies
starting from a modulator design with twice the \ac{OSR} and the square root of
$\gamma$, improves the noise floor by about $\unit[5]{dB}$ at any order. It is
worth underlining that this advantage does not reflect on the SNR. In fact,
since ΔΣ modulators have a strict limit on the input signal range that they can
tolerate, superimposing two signals at the input of the modulator requires
halving the maximum acceptable signal level. This lowers the maximum signal
power by $\unit[6]{dB}$. Hence, all together, one can expect an almost
unchanged SNR (in fact, almost negligibly degraded by about $\unit[1]{dB}$).

\section{Simulation and performance validation}
\label{sec:simulation}
For validation, the following test setup is considered. The two channels to be
simultaneously encoded have a \unit[20]{kHz} bandwidth and the overall \ac{OSR}
is set to 64. This sets the sample clock of the multiplexed system at
\unit[5.12]{MHz}. Binary modulators with $\STF(z)=1$ are considered. Test tones
with frequencies set at $\sim 1$ and $\unit[\sim 3.2]{kHz}$ and nominal
amplitudes set at $0.2$ and $0.44$ are used for the 1\Us{st} and 2\Us{nd}
channel respectively (amplitudes normalized with respect to the quantization
levels $\pm 1$). Merit factors include the quantization noise floor, SNR,
cross-talk, and maximum tolerable signal amplitude. The latter is evaluated by
forcing the same amplitude on the two signals and rising it to the point where
the modulator starts misbehaving. SNR is evaluated at the nominal and maximum
amplitudes, in band. A benchmark obtained using two individual ΔΣ coders for
the two signals is also provided. The benchmark coders have a \unit[20]{kHz}
bandwidth, half the order, and are designed with the same technique (Schreiers'
\texttt{synthesizeNTF} \cite{Schreier:UDSDC-2004}) and \ac{OSR} as those under
test, getting a \unit[2.56]{MHz} sample rate. All the tests can be replicated
using the PyDSM toolbox, available for download at
\url{http://pydsm.googlecode.com}.

\begin{figure}[t]
  \vspace{-1.5ex}
  \begin{tightcenter}
    \subfloat[\label{sfig:delsig-ntf-log}]{%
      \shortstack{\includegraphics[scale=0.54]{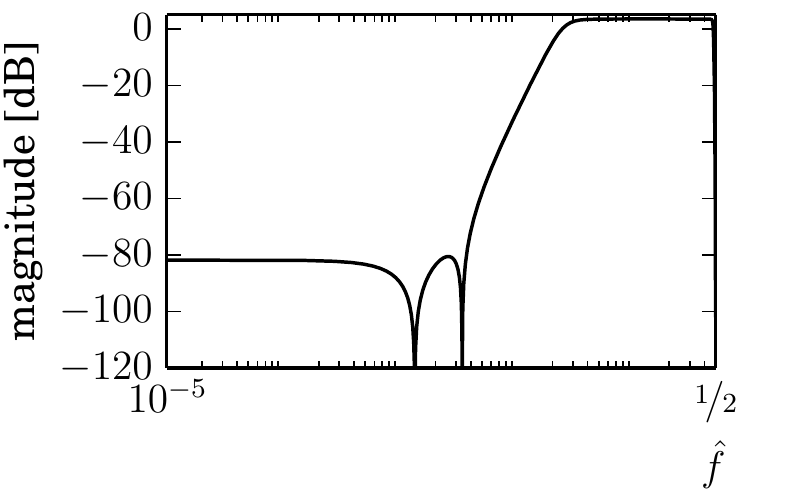}\\[-3ex]}}\hfill
    \subfloat[\label{sfig:delsig-ntf-lin}]{%
      \shortstack{\includegraphics[scale=0.54]{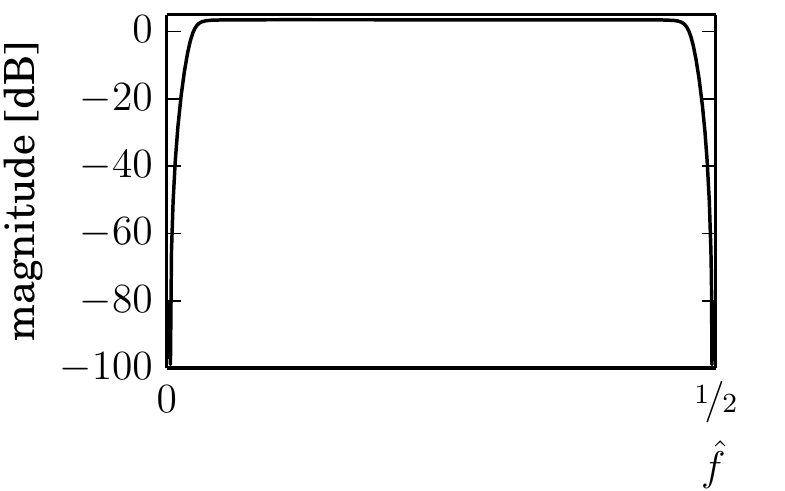}\\[-3ex]}}%
    \\[-1.5ex]
    \subfloat[\label{sfig:delsig-fragment}]{%
      \shortstack{\includegraphics[scale=0.54]{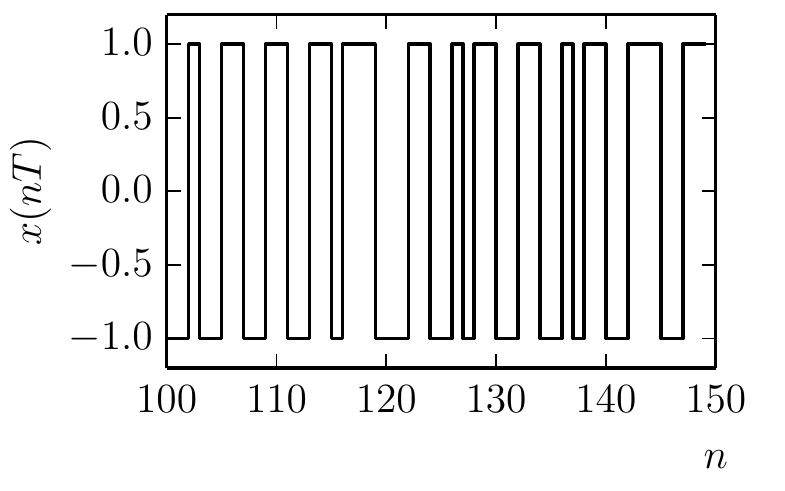}\\[-3ex]}}\hfill
    \subfloat[\label{sfig:delsig-pds}]{%
      \shortstack{\includegraphics[scale=0.54]{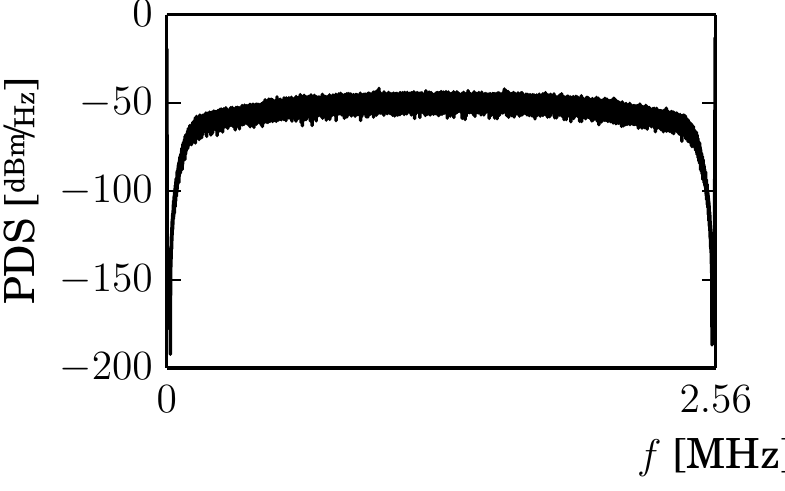}\\[-3ex]}}%
    \\[-1.5ex]
    \subfloat[\label{sfig:delsig-1000}]{%
      \shortstack{\includegraphics[scale=0.54]{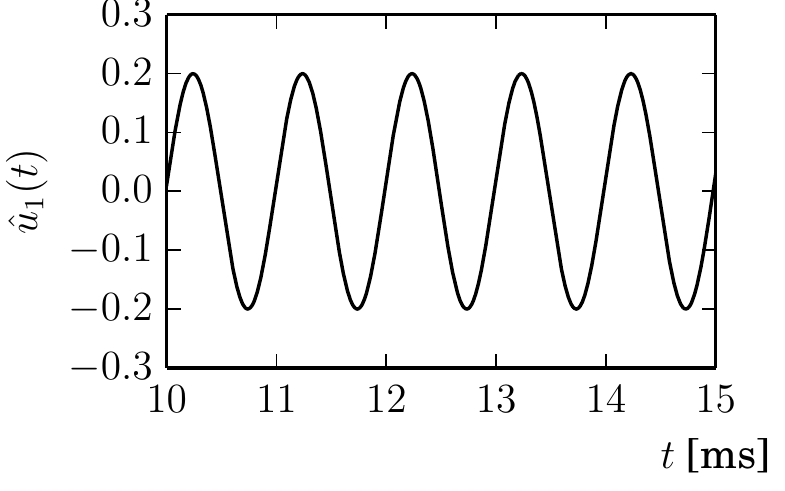}\\[-3ex]}}\hfill
    \subfloat[\label{sfig:delsig-3200}]{%
      \shortstack{\includegraphics[scale=0.54]{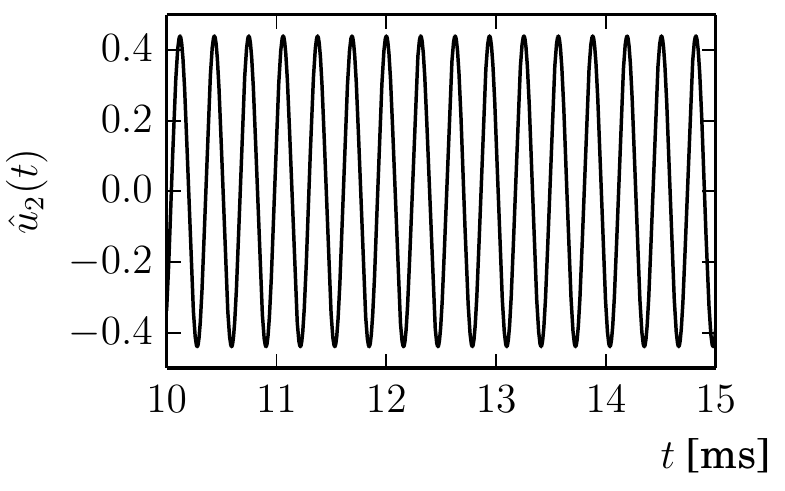}\\[-3ex]}}%
    \\[-1.5ex]
    \subfloat[\label{sfig:delsig-qnoise1}]{%
      \shortstack{\includegraphics[scale=0.54]{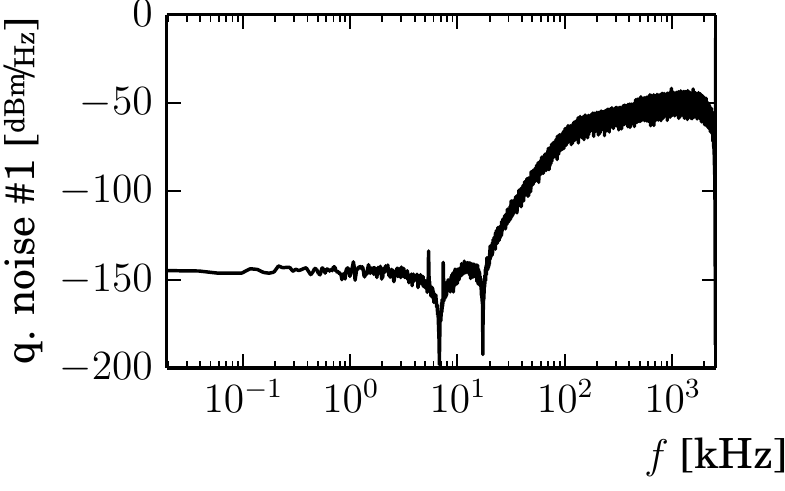}\\[-3ex]}}\hfill
    \subfloat[\label{sfig:delsig-qnoise2}]{%
      \shortstack{\includegraphics[scale=0.54]{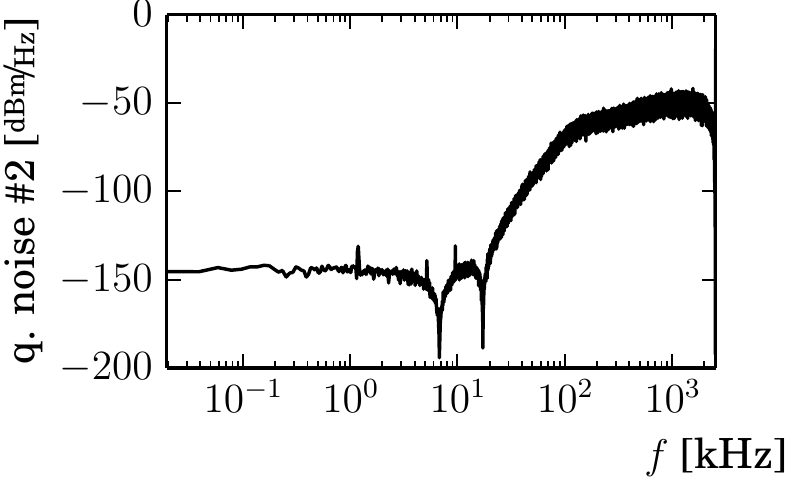}\\[-3ex]}}%
  \end{tightcenter}
  \caption{Behavior of dual channel ΔΣ modulator in the test case for an
    8\Us{th} order modulator designed by conventional
    techniques. In~\protect\subref{sfig:delsig-ntf-log}
    and~\protect\subref{sfig:delsig-ntf-lin}, \ac{NTF} magnitude response with
    log and linear frequency axis. In~\protect\subref{sfig:delsig-fragment},
    fragment of modulator output
    $x(nT)$. In~\protect\subref{sfig:delsig-fragment}, \ac{PSD} of
    $x(nT)$. In~\protect\subref{sfig:delsig-1000}
    and~\protect\subref{sfig:delsig-3200}, fragment of the reconstructed output
    signals $\hat u_1(t)$ and $\hat
    u_2(t)$. In~\protect\subref{sfig:delsig-qnoise1}
    and~\protect\subref{sfig:delsig-qnoise2}, \ac{PSD} of the quantization
    noise for the 1\Us{st} channel $x(nT)$ and the downconverted 2\Us{nd}
    channel $y(nT)=x(nT)\cdot(-1)^n$ .}
  \label{fig:delsig}
\end{figure}

The achieved behavior is illustrated in
Fig.~\ref{fig:delsig}. Plots~\subref{sfig:delsig-ntf-log}
and~\subref{sfig:delsig-ntf-lin} show the \ac{NTF} magnitude response with a
log and linear frequency axis respectively. Plot~\subref{sfig:delsig-fragment}
is a fragment of the modulator output $x(nT)$. Plot~\subref{sfig:delsig-pds}
illustrates the \ac{PSD} of $x(nT)$, obtained from time domain
simulations. Here, the overall shape of the noise \ac{PSD} agrees with the
curve in~\subref{sfig:delsig-ntf-lin}, while the peaks corresponding to the
test tones are almost invisible, being quite close to the plot
frame. Plots~\subref{sfig:delsig-1000} and~\subref{sfig:delsig-3200} show
fragments of the reconstructed signals $\hat u_1(t)$ and $\hat u_2(t)$. This
visually illustrates that the approach works and that there is no distortion,
noise, or cross-talk perceivable ``by the eye''. Eventually,
plots~\subref{sfig:delsig-qnoise1} and~\subref{sfig:delsig-qnoise2} show the
\ac{PSD} of the quantization noise for the modulator output $x(nT)$ (from which
the first output is obtained) and for $x(nT)\cdot(-1)^n$ (from which the second
output is obtained).  Quantitatively, performance is summarized in
Tbl.~\ref{tbl:delsig-performance}, that includes comparison to a reference
system.  For the multiplexed arrangement, he tabled maximum input amplitude is
cumulative. Namely, if both channels are active, each must be limited to half
the max cumulative value. Maximum SNR is reported with respect to this
situation.

\begin{table}[t]
  \caption{Performance indicators for a modulator system for stereo channels
    designed by the proposed method and conventional (reference) techniques}
  \label{tbl:delsig-performance}
  \begin{tightcenter}
    \begin{tabular}{lrr}
      \toprule
      \multicolumn{1}{p{4cm}}{8\Us{th} order dual channel modulator}
      & channel \#1& channel \#2\\
      \midrule
      Noise floor (in band) &\unit[-102]{dBm}& \unit[-101]{dBm} \\
      SNR (in band) & \unit[98]{dB} & \unit[105]{dB} \\
      Max SNR (in band) & \unit[103]{dB} & \unit[103]{dB} \\
      Crosstalk & \multicolumn{2}{c}{Below noise floor}\\
      Max input amplitude (cumulative) &
      \multicolumn{2}{c}{0.68}\\ 
      \midrule
      \multicolumn{1}{p{4cm}}{Reference 4\Us{th} order modulator}
      & test signal \#1 & test signal \#2\\
      \midrule
      Noise floor (in band) &\unit[-99]{dBm}& \unit[-98]{dBm} \\
      SNR (in band) & \unit[95]{dB} & \unit[101]{dB} \\
      Max SNR (in band) & \unit[68]{dB} & \unit[75]{dB} \\
      Crosstalk & \multicolumn{2}{c}{Below noise floor}\\
      Max input amplitude & 0.64 & 0.64 \\
      \bottomrule
    \end{tabular}
  \end{tightcenter}
\end{table}

\section{Conclusions}
\label{sec:conclusions}
A method for multiplexing stereo signals on a single \ac{DDSM} has been
proposed. Simulation data confirms the theoretical expectation that, with
respect to a reference system using two \acp{DDSM}, noise floor is improved
(\unit[-4]{dBm}) while SNR is degraded in an almost inappreciable way (-1 to
\unit[-2]{dB}). With respect to the reference system, the approach saves 1
modulator and 1 data link. Yet, the saving on the modulator is relative, since
the order of its filters is doubled. Note that the data link saving could in
principle be achieved also using two modulators and a conventional digital
multiplexing system. However, in the proposed arrangement demultiplexing is
simpler and only requires hardware on a single channel.  Most important, the
proposed approach is more flexible than a two-\ac{DDSM} one. In fact, if one
channel is unused, it lets the input range of the other be risen, so that for
the used channel the noise floor improvement can be capitalized into an SNR
improvement (up to \unit[+4]{dB}).

\bibliographystyle{SC-IEEEtran}
\bibliography{macros,IEEEabrv,various,sensors,analog,chaos}

\begin{thebibliography}{1}
\providecommand{\doi}[1]{DOI:#1}
\providecommand{\url}[1]{#1}
\csname url@samestyle\endcsname
\providecommand{\newblock}{\relax}
\providecommand{\bibinfo}[2]{#2}
\providecommand{\BIBentrySTDinterwordspacing}{\spaceskip=0pt\relax}
\providecommand{\BIBentryALTinterwordstretchfactor}{4}
\providecommand{\BIBentryALTinterwordspacing}{\spaceskip=\fontdimen2\font plus
\BIBentryALTinterwordstretchfactor\fontdimen3\font minus
  \fontdimen4\font\relax}
\providecommand{\BIBforeignlanguage}[2]{{%
\expandafter\ifx\csname l@#1\endcsname\relax
\typeout{** WARNING: IEEEtran.bst: No hyphenation pattern has been}%
\typeout{** loaded for the language `#1'. Using the pattern for}%
\typeout{** the default language instead.}%
\else
\language=\csname l@#1\endcsname
\fi
#2}}
\providecommand{\BIBdecl}{\relax}
\BIBdecl

\bibitem{Pamarti:TCAS1-54-3}
S.~Pamarti, J.~Welz, and I.~Galton, ``Statistics of the quantization noise in
  1-bit dithered single-quantizer digital delta-sigma modulators,''
  \emph{{IEEE} Trans. Circuits Syst. {I}}, vol.~54, no.~3, pp. 492--503, Mar.
  2006.

\bibitem{Harris:WET-2003}
F.~Harris, ``Sigma-delta converters in communication systems,'' in \emph{Wiley
  Encyclopedia of Telecommunications}, J.~G. Proakis, Ed.\hskip 1em plus 0.5em
  minus 0.4em\relax John Wiley \& Sons, Inc., 2003, vol.~IV, pp. 2227--2247.

\bibitem{Bizzarri:ISCAS-2012}
F.~Bizzarri, S.~Callegari, and G.~Gruosso, ``Towards a nearly optimal synthesis
  of power bridge commands in the driving of {AC} motors,'' in
  \emph{Proceedings of ISCAS 2012}, Seoul, May 2012, pp. 2119--2122.

\bibitem{Dunn:JAES-45-4}
C.~Dunn and M.~Sandler, ``Psychoacoustically optimal sigma delta modulation,''
  \emph{Journal of the Audio Engineering Society (AES)}, vol.~45, no.~4, pp.
  212--223, Apr. 1997.

\bibitem{Callegari:TCAS2-2013}
S.~Callegari and F.~Bizzarri, ``Noise weighting in the design of ΔΣ
  modulators (with a psychoacoustic coder as an example),'' \emph{{IEEE} Trans.
  Circuits Syst. {II}}, 2013, in press.

\bibitem{Schreier:UDSDC-2004}
R.~Schreier and G.~C. Temes, \emph{Understanding Delta-Sigma Data
  Converters}.\hskip 1em plus 0.5em minus 0.4em\relax Wiley-IEEE Press, 2004.

\bibitem{Callegari:TCAS1-2013}
S.~Callegari and F.~Bizzarri, ``Output filter aware optimization of the noise
  shaping properties of ΔΣ modulators via semi-definite programming,''
  \emph{{IEEE} Trans. Circuits Syst. {I} \emph{in press}}, 2013.
  \doi{10.1109/TCSI.2013.2239091}

\bibitem{Lee:Thesis-1987}
W.~L. Lee, ``A novel high order interpolative modulator topology for high
  resolution oversampling {A/D} converters,'' Master's thesis, Massachussets
  Institute of Technology, 1987.

\end{thebibliography}

\end{document}